# Steady-state cooperative emission of interlayer excitons in atomically thin light-emitting diodes

Andrés M. Mier Valdivia[1], Pavel E. Dolgirev[2], Grace H. Chen[1], Nadine Leisgang[2,3,4], Andrew Y. Joe[2,5], Stefan Ostermann[2], Deepankur Thureja[6], Yongxin Zeng[7], Dapeng Ding[2], Daniel Rhodes[8,9], Bumho Kim[9,10], Song Liu[9], Takashi Taniguchi[11], Kenji Watanabe[12], James C. Hone[9], Susanne F. Yelin[2], Hongkun Park[2,6], Mikhail D. Lukin[2], and Philip Kim[1,2,*]

[1] John A. Paulson School of Engineering and Applied Sciences, Harvard University, Cambridge, MA 02138, USA.

[2] Department of Physics, Harvard University, Cambridge, MA 02138, USA.

[3] Department of Physics, Philipps-University Marburg, 35032 Marburg, Germany.

[4] mar.quest | Marburg Center for Quantum and Sustainable Technologies, Phillipps-University Marburg, 35032 Marburg, Germany.

[5] Department of Physics and Astronomy, University of California, Riverside, CA 92521, USA.

[6] Department of Chemistry and Chemical Biology, Harvard University, Cambridge, MA 02138, USA.

[7] Department of Physics, Columbia University, New York, NY 10027, USA.

[8] Department of Materials Science and Engineering, University of Wisconsin, Madison, WI 53706, USA.

[9] Department of Mechanical Engineering, Columbia University, New York, NY 10027, USA.

[10] Department of Physics and Astronomy, University of Pennsylvania, Philadelphia, PA 19104, USA.

[11] Research Center for Materials Nanoarchitectonics, National Institute for Materials Science, 1-1 Namiki, Tsukuba 305-0044, Japan.

[12] Research Center for Electronic and Optical Materials, National Institute for Materials Science, 1-1 Namiki, Tsukuba 305-0044, Japan.

*Corresponding author: pkim@physics.harvard.edu

When an ensemble of emitters is confined within an optical wavelength, its radiation can become correlated, leading to cooperative emission. In two-dimensional (2D) transition metal dichalcogenide (TMD) electron-hole bilayers, strong light-matter coupling enables the formation of interlayer excitons (IXs), long-lived dipolar excitations that provide an ideal setting for emergent collective phenomena. Here, we engineer an atomically thin TMD light-emitting diode (LED) that generates electroluminescence (EL) from electrically injected IXs that are confined near the sample edges by an edge-enhanced fringing electric field. By measuring Hanbury-Brown-Twiss photon-photon correlations, we find pronounced signatures of collective decay in the form of large photon bunching followed by finite delay antibunching and bunching revivals. These emitter correlation signatures disappear when free charge carriers are doped to form a degenerate Fermi sea but are enhanced by the introduction of localized electrons. The results are consistent with a many-body quantum master equation that models the steady-state superradiant photon statistics of incoherently injected IXs. Our work establishes steady-state cooperative emission in a 2D TMD LED as a platform for studying strong light–matter interactions in electrically driven, non-equilibrium phases of Bose-Fermi mixtures.

The strong coupling between light and matter can generate quantum collective effects that lead to cooperative emission, where photon-mediated interactions among emitters drive the buildup of correlations (*1*, *2*). This collective behavior leads to correlated photon bunching in the emitted radiation. R. H. Dicke initially formulated the theory of superradiance in 1954 (*1*) , and the effect was subsequently first observed in 1973 in a molecular gas (*3*). Since then, the observation of superradiant effects has expanded into material platforms to include molecules embedded in crystals (*4*), molecular aggregates (*5*), quantum dots (*6*, *7*) and perovskites (*8*). Beyond its fundamental value, superradiance holds technological promise for developing thresholdless lasers (*9*, *10*).

In the solid-state, optically excited emitters couple strongly to the environment, causing them to undergo rapid decoherence. Consequently, observing superradiant phenomena often requires careful selection of systems whose properties can be tightly controlled to form a homogeneous ensemble of two-level emitters (*5*, *11*), or by imposing favorable conditions with strong magnetic fields or femtosecond laser excitation (*8*, *12*). Another approach to search for materials that can sustain cooperative light emission in the fast decoherence limit is to combine strongly interacting quasiparticles with device engineering (*11*, *13*, *14*).

Interlayer excitons (IXs) in type-II two-dimensional (2D) semiconductor heterostructures present a compelling system for exploring strong light-matter coupling owing to the spatial separation of electrons and holes in different layers (Fig. 1A,B). In $MoSe_2/WSe_2$ heterobilayers, IXs boast lifetimes approaching a microsecond (*15*), orders of magnitude longer than intralayer excitons (*16*). Their out-of-plane electric dipole allows precise electrostatic tunability of the emission energy (*15*, *17*), facilitating the realization of tunable IX trap devices (*18*, *19*) and 2D excitonic transistors (*20*). Given that IXs are composite bosons, the ongoing search for an IX Bose-Einstein condensate (BEC) ground state (*21*, *22*) has driven the development of measurement techniques to probe excitonic coherence (*23*–*26*). An open critical question is how light-mediated correlations reshape this rich phase diagram. Further, we are motivated by the prospect of using the nonlinear optical properties of 2D semiconductors to develop sources of nonclassical light (*13*) or coherent photons for atomically thin nanoscale lasers (*27*).

The $MoSe_2/WSe_2$ heterostructure offers the additional advantage of forming an atomically thin light-emitting diode (LED), enabling electrical injection and electroluminescence (EL) of IXs with a voltage bias (*15*, *25*, *28*, *29*). Photoluminescence (PL), used in most IX studies, presents several inherent limitations that potentially hinder the observation of photon-mediated correlations compared to EL. First, due to their weak oscillator strength, IXs are typically excited using a far off-resonance laser, initially generating intralayer electron-hole (*e*-*h*) pairs that relax into the interlayer state (*30*). Second, a coherent excitation can imprint its phase onto the IXs, complicating the unambiguous identification of spontaneously developing correlations. Third, the excitation area in far-field optics is limited by diffraction to $\sim 1\ \mu m^2$. Electrical injection of excitons in our LED overcomes these constraints: electrons and holes are incoherently injected at the quasi-Fermi levels, enabling the direct formation of IXs from uncorrelated *e*-*h* pairs. In addition, engineered device design and electrostatic gate control set the recombination region rather than the excitation optics (*31*, *32*).

Here, we report the observation of steady-state cooperative EL from incoherently injected IXs in atomically thin $MoSe_2/WSe_2$ LEDs. Our engineered devices enables sub-diffraction confinement of IXs, readily satisfying the Dicke criterion for superradiance (*1*). We employ Hanbury-Brown-Twiss (HBT) interferometry to quantify the second-order correlation function $g^{(2)}(\tau)$ of the emitted light field at a finite delay time $\tau$ (*33*). We observe signatures of DC current driven cooperative IX emission by measuring pronounced photon bunching ($g^{(2)}(0) > 2$) followed by oscillatory dynamics exhibiting antibunching ($g^{(2)}(\tau) < 1$) and subsequent bunching revivals ($g^{(2)}(\tau) > 1$). Doping free carriers pushes the system out of this cooperative state due to the interaction of IXs with a delocalized, degenerate Fermi sea. We develop a quantum master equation to explain that the observed features arise due to steady-state superradiant photon statistics and to identify that the incoherent injection of *e*-*h* pairs is an important ingredient underlying the large emergent collective decay.

We fabricate a dual gated $MoSe_2/WSe_2$ device with a monolayer hexagonal boron nitride (h-BN) spacer selectively inserted to separate part of the heterobilayer as shown in Fig. 1A, as described in the Materials and Methods (*34*). The TMD monolayers are angle-aligned near 0° to produce a type-II band gap at the *K*/-*K* points of the Brillouin zone. The h-BN dielectric spacer restricts direct interlayer coupling to a ∼ 1 $\mu$m-wide strip near the $WSe_2$ flake edge (Fig. S1). Each TMD monolayer is independently contacted with bottom platinum electrodes with corresponding contact gates. The contact gates locally dope the regions beneath, creating separate electron and hole reservoirs that facilitate carrier injection into the heterobilayer *p*-*n* junction region of the device (*35*, *36*). All measurements are performed at a temperature $T = 6$ K unless otherwise stated.

In equilibrium, as shown in Fig. 1B, the electron and hole chemical potentials, $\mu_e$ and $\mu_h$, are spatially uniform. A forward bias $-V_b$ on $MoSe_2$ while keeping $WSe_2$ grounded splits $\mu_e$ and $\mu_h$ asymmetrically such that only electrons are injected, resulting in no IX EL (Fig. 1C) and weak interlayer current $I$ (< 10 nA), as shown in Fig. 1D. To enable hole injection, we symmetrically shift the chemical potentials by applying scaled top and bottom gate voltages, $V_{tg} = V_g$ and $V_{bg} = \alpha V_g$, where $\alpha$ is the gate efficiency ratio defined by dividing the ratio of the bottom-over-top h-BN dielectric layer thicknesses. At $V_g \sim -0.65$ V, the sharp onset of IX emission is accompanied by a pronounced increase in the current (Fig. 1C-D). This marked change identifies the gate configuration where $\mu_h$ reaches the top of the $WSe_2$ valence band, enabling hole injection. Beyond this threshold, EL and $I$ exhibit non-monotonic behavior due to changes in the charge configuration of the IX states. This gate and bias controlled electrostatic configuration of the 2D *p*-*n* junction is also confirmed by $V_g$ and $V_b$ dependent reflectance spectrum maps in the heterobilayer region (Section 1 of (*34*)).

IX EL appears in only three diffraction-limited spots along the edge of the heterobilayer's overlap region defined by the $WSe_2$ flake boundary (Fig. 1E). In this area, the geometric discontinuity leads to an abrupt change in the charge density that generates a locally enhanced electric field near the edge. This fringing electric field lensing causes electrons in $MoSe_2$ and holes in $WSe_2$ to accumulate, leading to preferential recombination at the edge of the heterobilayer (*32*, *37*). Pronounced photocurrent along this edge (Fig. 1F) from intralayer *e-h* dissociation (*38*) and our finite element method-based simulations (Section 2 of (*34*)) support the nanometer-scale confinement of IXs. Below, we focus on the EL from the left corner of the heterostructure (white-dashed circle in Fig. 1E-F). Measurements in the other two bright spots and on additional devices yield consistent results (Sections 3 and 4, respectively, of (*34*)).

Fig. 2A shows gate dependent EL spectrum at a representative fixed bias voltage $V_b$ = 2.8 V. We simultaneously monitor $I$ to identify transport signatures of charge transitions (Fig. 2B). To further characterize the system, we perform PL spectroscopy of both interlayer excitons and $MoSe_2$ intralayer excitons, as well as reflectance contrast $\Delta R/R_0$ spectroscopy, which has been previously used to map the doping configuration in h-BN separated TMD heterobilayers (*39*). From these combined characterization methods (Section 5 of (*34*)), we find four distinct IX charge states: (i) $IX^0$: charge neutrality, characterized by dominant IX EL above 1.35 eV; (ii) $IX^+$: positively charged IX; (iii) $IX^-_{IGS}$: weakly localized negatively charged IX interacting with electrons captured by in-gap states (IGS) in $MoSe_2$; and (iv) $IX^-$: negatively charged IX with degenerate *n*-doping.

Compared to charge neutrality, both *n*- and *p*-type degenerately doped regimes exhibit charged $IX^\pm$ EL dominated by emission below 1.35 eV. This peak energy shift is driven by a gate-induced *e-h* imbalance, as reported in previous work (*15*, *29*). In contrast, the EL from $IX^0$ and $IX^-_{IGS}$ shows gradual intensity variations while remaining at constant energies. In particular, the EL intensity of the peaks above 1.35 eV decreases as the system approaches $IX^-$. The strong $MoSe_2$ charged exciton PL, together with nearly unchanged neutral $MoSe_2$ exciton $\Delta R/R_0$, indicates the presence of doped electrons that do not form a Fermi sea (Fig. S9). This interpretation is further supported by reports of IGS in flux-grown $MoSe_2$ crystals, resulting in localized, defect-bound trions (*40*). We note that the positively charged IX does not exhibit an appreciable localized regime in our device, possibly because of the higher mobility in the $WSe_2$ crystals used in the devices (*41*).

The precise electrical control over the relative *e-h* densities reveals a rich phase space of EL IX states. Motivated by the evidence of confined emitters, we investigate the EL photon statistics using a fiber-based Hanbury-Brown-Twiss interferometer (*33*) (Materials and Methods (*34*)) to determine whether cooperative excitonic states are present in our device. This setup measures photon coincidences between two detectors and thereby probes the second order correlation function $g^{(2)}(\tau)$, where $\tau$ is the photon delay time between photon arrivals of the two detection arms. The value of $g^{(2)}(0)$ is sensitive to the statistical properties of the emitters' quantum state (*42*): $g^{(2)}(0) = 1$ indicates uncorrelated, Poissonian emission; $g^{(2)}(0) > 1$ indicates photon bunching arising from excess correlations between emitters, e.g. a signature of cooperative emission; and $g^{(2)}(0) < 1$, or antibunching, indicates a tendency towards discretely spaced emission events with $g^{(2)}(0) = 0$ corresponding to the ideal single photon emission limit.

Fig. 2C-F show representative $g^{(2)}(\tau)$ curves measured in the regimes corresponding to $IX^+$, $IX^0$, $IX^-_{IGS}$, and $IX^-$, respectively. While $g^{(2)}(\tau) \approx 1$ for the EL from $IX^\pm$ indicates uncorrelated emission in this regime, we observe bunching photon statistics for $IX^0$ and $IX^-_{IGS}$, where $g^{(2)}(0) > 1$ and decays with a characteristic decay time $\tau_d$. For further quantitative analysis, we plot $g^{(2)}(0)$ and $\tau_d$ as a function of $V_g$ in Fig. 2G (see Section 6 of (*34*) for fitting procedure). Here, starting from the net *p*-doping regime ($V_g < -1.10$ V), $IX^+$ exhibits uncorrelated emission, since $g^{(2)}(\tau) = 1$ for all $\tau$ (Fig. 2C). Upon entering charge neutrality ($-1.10 < V_g < -0.95$ V), $IX^0$ develops a pronounced bunching peak $g^{(2)}(0) > 1$ (Fig. 2D), whose magnitude

increases towards $g^{(2)}(0) \sim 2$ at more negative $V_g$. The decay time $\tau_d$ is between 100 − 300 ns, consistent with the measured IX PL lifetime (Fig. S11). At $V_g = -0.95$ V, localized electrons enter the heterobilayer and bunching becomes remarkably enhanced, increasing beyond $g^{(2)}(0) = 2$, the thermal bunching limit. Even more intriguingly, $g^{(2)}(\tau)$ develops dynamics characterized by finite-delay antibunching followed by a subsequent bunching revival (Fig. 2E). In this regime, the fast Fourier transform (FFT) of $g^{(2)}(\tau)$ develops a peak at finite frequency, consistent with the observed oscillatory behavior (Fig. 2H). We define the inverse of this frequency as the oscillation period $\tau_{osc}$, which describes the finite delay dynamics (Fig. 2I). We find that $\tau_{osc}$ ranges from 1 − 5 $\mu$s, more than an order of magnitude longer than $\tau_d \sim 50 - 150$ ns. At $V_g = -0.85$ V, the bunching correlations become most pronounced: $g^{(2)}(0) \sim 5$, the finite time antibunching dip approaches ∼ 0.70, and the bunching revival reaches an amplitude of ∼ 1.05 (Fig. 2E). Further increasing the gate voltage progressively weakens these correlations until they vanish entirely when an electron Fermi sea forms at $V_g = -0.70$ V (Fig. 2F).

The observed photon statistics in the $IX^0$ and $IX^-_{IGS}$ regimes are consistent with cooperative emission of electrically injected IXs. Similar to Dicke superradiance in atomic systems and quantum dots (*11*), this phenomenon emerges when strong light-matter coupling makes the photon emitters indistinguishable within a volume, $V$, smaller than the volume set by the emitted photon's wavelength, $\lambda$, i.e., $V \lesssim \lambda^3$ (*1*, *2*, *5*). In most 2D excitonic systems, $\lambda$ ranges between ∼0.5−1 $\mu$m, so this geometric condition is readily satisfied at typical optical excitation densities of $10^{10}$ to $10^{12}$ $cm^{-2}$, corresponding to an average interparticle spacing of ∼ 10 − 100 nm. Despite emerging work studying collective emission in moiré heterobilayers (*14*), realizing correlated decay is challenging due to several factors. First, excitons in TMD semiconductors have Bohr radii on the order of a few nanometers (*43*) so their wavefunction extends over many atomic unit cells. This relatively large exciton extent makes TMD excitons highly susceptible to decoherence arising from disorder and phonon scattering (*11*). Second, structural inhomogeneity in heterobilayers due to van der Waals assembly introduces additional spectral broadening within a diffraction-limited spot (*44*). Third, laser excitation is usually off-resonant and populates all available low energy optical states within a ~ 1 $\mu m^2$ region, including free IXs that can diffuse away from the excitation spot (*15*). Together, these factors often inhibit the formation of collective optical correlations, consistent with the absence of photon bunching in IX PL even under similar doping conditions to the EL measurements (Fig. S12).

In our IX LEDs, device engineering facilitates the Dicke criterion by producing electric field-enhanced submicron confinement of IXs at the sample boundaries (Fig. 1E-F). Complementarily, electrical injection takes maximum advantage of these geometric aspects by overcoming the limitations of off-resonant diffraction-limited optics. An additional benefit of *e-h* recombination in LEDs is that it typically occurs through defect states due to the Shockley-Read-Hall (SRH) mechanism (*31*). This favors preferential recombination through the electronic $MoSe_2$ IGS, whose localization reduces the spatial extent of the IX wavefunction. The disappearance of correlations upon degenerately doping either electron or hole Fermi seas, where delocalized polarons are expected to form (*45*), further highlights the importance of localized rather than extended IX states for the cooperative emission.

To gain further insight into the origin of this excitonic collective decay, we perform spectrally resolved $g^{(2)}(\tau)$ measurements of $IX^{-}_{IGS}$ using tunable filters. The full EL spectrum exhibits strong bunching ($g^{(2)}(0) > 2$) together with the finite-delay dynamics described above (Fig. 3A). We then measure $g^{(2)}(\tau)$ within narrower energy bands defined by the color-shaded regions in Fig. 3A. The low energy peaks of the spectrum (< 1.33 eV) show weak or negligible correlations (Fig. 3B-C). The band capturing the ~ 1.35 eV peaks exhibits correlations that are enhanced relative to the full spectrum (Fig. 3D), suggesting that this energy band range is the primary source of the observed non-trivial $g^{(2)}(\tau)$. The high-energy tail (> 1.36 eV) also exhibits finite bunching (Fig. 3E), although we cannot rule out possible contributions from spectral overlap with the tail distribution of the neighboring peaks. We attribute the lower energy spectrum (< 1.33 eV) to deeply trapped IXs and conclude that finite photon-photon correlations arise only from $IX^0$ and $IX^-_{IGS}$.

The temperature dependence of $g^{(2)}(\tau)$ reveals the energy scales associated with the observed cooperative emission. Figure 3F shows $g^{(2)}(0)$ for $IX^0$ ($V_g = -0.95$ V) and $IX^-_{IGS}$ ($V_g = -0.85$ V) from 6 K to 80 K. In both regimes, correlations disappear between 70−80 K, corresponding to an energy of ~ 6−7 meV. For $T \leq$ 70 K, the EL spectra remain qualitatively similar to those measured at 6 K, whereas the $IX^0$ and $IX^-_{IGS}$ features broaden substantially by 80 K (Section 7 of (*34*)). Meanwhile, the current increases steadily with temperature, suggesting that changes in the transport alone cannot fully account for the non-monotonic temperature evolution of the bunching strength.

In our physical picture, a photoemission event depopulates the weakly localized excitonic states, which are subsequently replenished through incoherent recharging by injected carriers from the contact-gated regions of the device. Considering the device parameters, we estimate a slow recharging timescale of ~ 2 $\mu$s (Section 8 of (*34*)), offering a natural explanation for the long timescale associated with the minimum in the $g^{(2)}(\tau)$ trace in the IGS regime (Fig. 2E). In typical solid-state superradiant experiments, well-isolated emitters are photoexcited by a pulsed laser, closely resembling Dicke's original formulation where the individual two-level emitters in an ensemble are initially prepared to all be in the excited level (*1*, *2*, *5*, *11*). Even in past work on steady-state superradiance, the emitters are replenished via coherent optical excitation (*46*, *47*). In contrast, here the observed steady-state EL cooperative emission emerges from incoherent pumping, similar to sub-threshold semiconductor lasers. Unlike conventional lasers, however, our device lacks an optical cavity, representing a previously unexplored regime.

A previous TMD heterostructure experiment interpreted the emergence of photon bunching in intralayer EL as evidence for the condensate transition of IXs, where excess correlations arise from critical fluctuations (*25*). However, several aspects of our data are inconsistent with this scenario. First, a condensate should form throughout the bulk of the device, which cannot readily explain why the emission is restricted to the corners and edges (Fig. 1E-F). Second, excitonic condensation, which involves the onset of interlayer charge coherence, should therefore be accompanied by an abrupt change in device current due to coherent tunneling, yet no such signature is observed. Third, a simple estimate of the degeneracy temperature further disfavors this scenario. From the blueshift of the IX peak versus bias (Section 9 of (*34*)), we estimate $n_{\mathrm{eh}} \lesssim 10^{12}\ \mathrm{cm}^{-2}$, which provides an upper bound $\sim$ 45 K for the IX degeneracy temperature (Section 10 of (*34*)). Thus, condensation alone cannot account for correlations that persist to 70 K. Another potential source of fluctuations is the Mott transition associated with excitonic ionization (*22*). However, this transition is expected to occur at $n_{\mathrm{eh}} \sim 2 - 3 \times 10^{12}\ \mathrm{cm}^{-2}$ (*18*, *48*), well above our estimated density range.

We also exclude the following other potential sources of photon bunching based on our observations. The sensitivity of correlations to doping could suggest quantum beating between attractive and repulsive polaron branches. However, the relevant timescales would be set by the inverse of the Fermi energy $\sim \hbar/E_{\mathrm{F}}$. In TMDs, accessible Fermi energies range from $E_{\mathrm{F}} \sim 0.01 - 10$ meV ($10^{10} - 10^{13}\ \mathrm{cm}^{-2}$), resulting in much shorter timescales of $0.01 - 10$ ps than those observed experimentally. Thermal bunching from intensity fluctuations in uncorrelated ensembles is another possible mechanism (*49*). However, the corresponding timescale for these correlations should scale as the inverse linewidth $\hbar/\delta E$. Even when considering the narrowest reported emission linewidth for IXs in the literature, $\delta E \sim 100\ \mu\mathrm{eV}$ (*50*), the longest possible decay time is $\sim$ 100 ps, still orders of magnitude shorter than in our experiment. Cathodoluminescence-like large photon bunching (*51*) likewise fails to explain our observations since here *e-h* injection occurs at the quasi-Fermi levels such that each excitation produces at most one excited IX.

Our interpretation of the observed cooperative exciton decay is further captured by a minimal phenomenological model based on a quantum master equation, illustrated schematically in Fig. 4A, with only three parameters: the number of emitters $N$, the single-emitter radiative lifetime $\tau_{IX}$, and the incoherent recharging time $\tau_{ir}$. In this framework (Section 11 of (*34*)), an ensemble of $N$ emitters undergoes collective radiative decay while being incoherently repopulated on a much slower timescale $\tau_{ir}$. For our experimental conditions, we estimate $(\tau_{ir}/\tau_{IX})^{-1} \sim (2\ \mu s/100\ ns)^{-1} \simeq 0.05$. The resulting theoretical $g^{(2)}(\tau)$ traces, such as in the middle inset of Fig. 4E, reproduce the key experimental features shown in Fig. 2E. In particular, the model captures: (i) the bunching value of $g^{(2)}(0)$, which can in principle become arbitrarily large (Fig. 4B); (ii) the initial decay to min[$g^{(2)}(\tau)$], with a minimum value consistent with the data (Fig. 4C, top); (iii) the subsequent overshoot above unity (Fig. 4C, bottom); and (iv) the final decay to $g^{(2)}(\tau) \to 1$ as $\tau \to \infty$ (Fig. 4E). Interestingly, the model reveals two qualitatively distinct regimes separated by $(\tau_{ir}/\tau_{IX})^{-1} \approx 0.2$. For $(\tau_{ir}/\tau_{IX})^{-1} \gtrsim 0.2$, the bunching peak $g^{(2)}(0)$ increases with $N$ and $g^{(2)}(\tau)$ lacks finite-delay oscillations. This behavior is consistent with Dicke's original superradiance picture that stronger collective effects arise at higher emitter density (*1*, *2*). In the experimentally relevant slow recharging regime $(\tau_{ir}/\tau_{IX})^{-1} \lesssim 0.2$, the model instead predicts that $g^{(2)}(0)$ increases as $N$ decreases and that $g^{(2)}(\tau)$ develops pronounced finite-delay dynamics in agreement with the experimental observations. This slow-recharging-rate regime reveals a mechanism absent in the standard Dicke superradiance: the uncorrelated recharging channel populates not only bright Dicke states but also subradiant dark states. Because these dark states couple only weakly to the ground state, they act as long-lived reservoirs (*47*) that intermittently feed the superradiant Dicke manifold within the timescale of $\tau_{ir}$, explaining the experimentally observed finite-delay features in $g^{(2)}(\tau)$ (see SM.XV).

Varying the current I at fixed $V_g$ provides experimental control over the incoherent recharging rate $\tau_{ir}$, where $(\tau_{ir}/\tau_{IX})^{-1} \propto I$. In the Fig. 4D insets, we plot $g^{(2)}(0)$ versus $I$ at fixed gate voltage ($V_g = -1.0$ V, $IX^0$ regime) to hold the charge imbalance and radiative lifetime $\tau_{IX}$ constant while tuning $V_b$. We observe that $g^{(2)}(0)$ steeply decreases as $I$ increases. Since I grows superlinearly with bias ($I \propto V_b^{5.3}$, Fig. S15), whereas the exciton density $n_{\mathrm{eh}}$ increases linearly with bias ($n_{\mathrm{eh}} \sim V_b$, Fig. S15), the variation in $g^{(2)}(0)$ is dominated by the recharging rate change rather than changes in $n_{\mathrm{eh}}$. As shown in Fig. 4E, our quantum master equation model with fixed $N$ = 25 reproduces this rapid decline in $g^{(2)}(0)$ with increasing $(\tau_{\mathrm{ir}}/\tau_{\mathrm{IX}})^{-1}$. A faster replenishment rate re-populates the ensemble during its cascade down the symmetric Dicke ladder, effectively resetting it into a less correlated state. The modeled $g^{(2)}(\tau)$ curves (Fig. 4E, insets) also capture the full temporal evolution from giant zero-delay bunching ($\tau_{\mathrm{ir}} > \tau_{\mathrm{IX}}$, left inset), to oscillatory correlations ($\tau_{\mathrm{ir}} \sim \tau_{\mathrm{IX}}$, middle inset), and ultimately to flat $g^{(2)}(\tau)$ ($\tau_{\mathrm{ir}} \ll \tau_{\mathrm{IX}}$, right inset). In the low current limit ($I \sim 1$ nA), $g^{(2)}(0)$ reaches ∼ 13, which is the highest value reported $g^{(2)}(0)$ in TMD systems without high-energy incoherent cascade processes (*51*).

A limitation of our model is that it simplifies the many-body Bose-Fermi mixture in the device to an ensemble of two-level emitters interacting only through the photonic field. While this approximation successfully captures the dependence of cooperative emission on the recharging rate and emitter number, this model neglects the Coulomb interactions between IXs and free carriers. As we noted above, charge doping to create either electron or hole Fermi seas completely suppresses emitter correlations, whereas adding unpaired localized electrons enhances the emitter correlations. These effects cannot be explained solely by changes in the recharging rate. For example, the current decreases upon doping delocalized electrons, which in our phenomenological picture should enhance photon bunching rather than suppress it. Our results therefore suggest that the microscopic nature of the charged IXs, whether they form delocalized polarons or localized trions, is critical for mediating collective light emission.

Finally, we note that the measured dynamical photon-photon correlations are strongly reminiscent of the photon statistics observed in semiconductor nanolasers across the lasing transition (*52*, *53*). In nanoscale lasers, the lasing threshold can be crossed with a relatively small number of photons ($\lesssim 100$) coupled to the lasing mode (*27*), and pronounced photon number fluctuations drive operation in a regime fundamentally distinct from conventional lasers (*52*). Developing mesoscale lasers based on 2D materials is a tantalizing prospect, particularly since our device exhibits large photon bunching and dynamical correlations even in the absence of an optical cavity. Our finding thus motivates future measurements of first-order photon field coherence ($g^{(1)}(\tau)$) to advance, for example, the development of cooperative, mirrorless lasers (*10*). Furthermore, fast time-resolved pulsed EL measurements, enabled by high-frequency device architecture, may elucidate the transient formation of the incoherently pumped cooperative state and uncover the dynamics of its pulsed collective emission.

In conclusion, we report steady-state cooperative EL from uncorrelated injected charges into atomically thin TMD heterostructure LEDs. By combining transport and optical spectroscopic measurements, we show that electrically injected IXs can enter a cooperative emission regime governed by incoherent recharging. The resulting charge-tunable photon bunching, finite-delay antibunching and revival dynamics, coupled with intriguing interplay with doped charge carriers, reveal a unique form of collective light–matter interaction beyond the conventional Dicke paradigm. Our demonstration of electrically controllable steady-state collective decay in a solid-state device platform opens new opportunities for merging quantum optics and many-body phenomena in tunable optoelectronic devices.

**Author contributions**

AMMV performed the experiments with GHC, NL and AYJ. AMMV and AYJ fabricated the devices. PED performed the theoretical modeling with assistance from SO. DT performed the COMSOL simulations. DD wrote the code for computing the second-order correlation curve from raw data. DR, BK, SL and JCH provided the flux-grown TMD crystals. TT and KW grew the h-BN crystals. YZ and SFY collaborated on discussions relating to the interpretation of the results. PK, MDL and HP supervised the project. AMMV, PED, GHC and PK wrote the manuscript with input from all authors.

**Acknowledgements**

PK and AMMV acknowledge support from ONR MURI N00014-21-1-2377. JCH acknowledge support from the US Army Research Office MURI, grant no. W911NF2120147. PK and MDL acknowledge support from NSF QLCI (QSI-2553619). MDL acknowledges support from the NSF, grant no. PHY 2608266. MDL and SFY acknowledge the NSF through the CUA PFC (PHY-2317134). SFY acknowledges the AFOSR through FA9550-24-1-0311. HP acknowledges support from the NSF Center for Ultracold Atoms (PHY-1125846). TT and KW acknowledge the support from the JSPS KAKENHI (Grant Nos. 21H05233 and 23H02052), the CREST (JPMJCR24A5), JST and World Premier International Research Center Initiative (WPI), MEXT, Japan. N.L. acknowledges support from the Swiss National Science Foundation (SNF) (Project No. P500PT_206917), the German Research Foundation (DFG) Emmy Noether Program (Project No. 553875273), the LOEWE Start Professorship Program of the State of Hesse (Az. III5 - 519/05/01.002 - (0021)), and the Wübben Stiftung Wissenschaft Appointment Accelerator Program (Az. 0159-2025-AA). Nanofabrication was performed at the Center for Nanoscale Systems at Harvard University, supported in part by an NSF NNIN award ECS-00335765. PED acknowledges fruitful discussions with P. Nosov and E. Demler. AMMV and PK acknowledge helpful discussions with C. Stephen Hellberg.

**Competing interests**: Authors declare that they have no competing interests

**Data, code, and materials availability**: Available upon reasonable request.

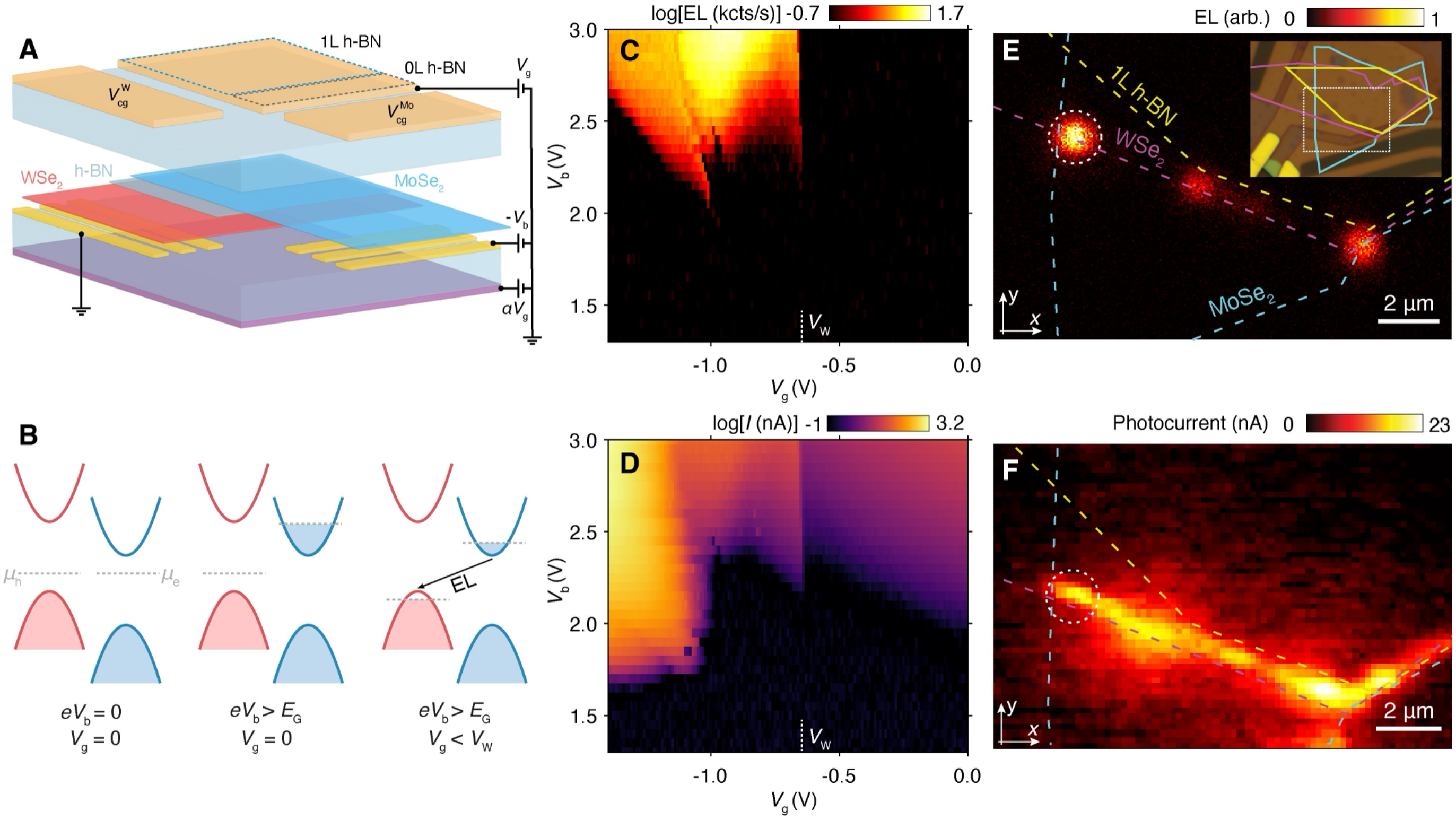

FIG. 1. **An atomically thin interlayer exciton light-emitting diode**. **A** Schematic of the dual-gated, h-BN encapsulated $WSe_2/MoSe_2$ heterobilayer device. The bottom and top h-BN thicknesses are 75 nm and 52 nm, respectively, setting the gate efficiency ratio to $\alpha = 75/52 \approx 1.442$. The $WSe_2$ and $MoSe_2$ contact gate voltages are kept fixed at $V_{cg}^{W} = -14.5$ V and $V_{cg}^{Mo} = 15$ V. **B** Schematic band diagrams of the heterobilayer under different bias ($V_b$) and gate ($V_g$) voltage conditions. $E_G$ denotes the type-II band gap between the $WSe_2$ valence band (red) and $MoSe_2$ conduction band (blue). A negatively applied bias voltage $eV_b$ on $MoSe_2$ can shift $\mu_e$ upwards, making the $MoSe_2$ layer become electron-doped when $eV_b > E_G$. Increasing $V_g$ shifts both $\mu_h$ and $\mu_e$ such that for nonzero $V_W \geq V_g$, the grounded $WSe_2$ layer simultaneously acquires hole doping. **C** The integrated IX EL intensity and **D** the current $I$ across the device as functions of $V_b$ and $V_g$. The discontinuity at $V_g = V_W$ marks the onset of hole injection and the electrical generation of IXs. **E** Spatial map of the integrated IX EL, showing emission localized primarily at the device edge and corners. Inset: optical micrograph of the device, with the dashed square indicating the region corresponding to panels E and F. **F** Scanning photocurrent map, where the signal indicates enhanced current when the laser is focused at a given position. The pronounced response along the heterobilayer edge indicates an electric field lensing effect near the $WSe_2$ flake boundary.

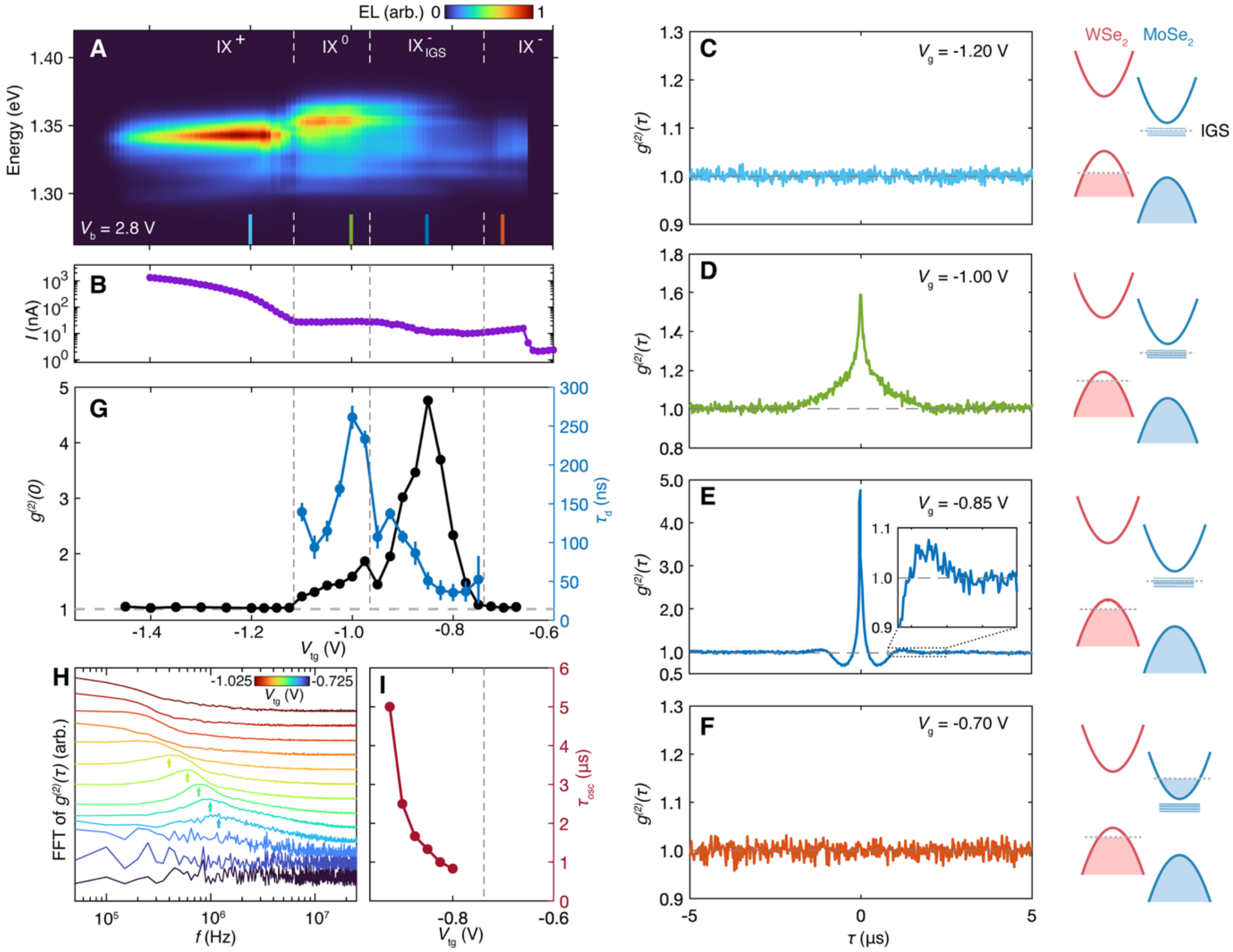

FIG. 2. **Evolution of photon–photon correlations across the doping regimes.** Four distinct doping regimes, representing EL optical species $IX^+$, $IX^0$, $IX^-_{IGS}$ and $IX^-$, separated by dashed vertical lines, emerge as $V_g$ is tuned at fixed bias $V_b = 2.8$ V, as seen in the EL spectra (**A**) and current $I$ (**B**). **C-F** Second-order photon correlation function $g^{(2)}(\tau)$ measured in each regime at $V_g = -1.20$, $-1.00$, $-0.85$, and $-0.70$ V, respectively. The curve colors correspond to the colored vertical line in (A). Next to each $g^{(2)}(\tau)$, and schematic band diagrams indicate the associated doping configuration. Pronounced photon bunching is observed in (D) and (E), with (E) exhibiting characteristic dynamics where $g^{(2)}(\tau)$ initially decays rapidly, reaches a minimum below unity on a timescale $\tau_{osc}$, and subsequently overshoots above unity (inset of E). **G** Zero-delay bunching amplitude $g^{(2)}(0)$ (black, left axis) and its decay time $\tau_d$ (blue, right axis) as a function of $V_g$. **H** The fast Fourier transform (FFT) of $g^{(2)}(\tau)$ shows the emergence of a peak (arrows) in the $IX^-_{IGS}$, regime, corresponding to dynamical correlations like the one shown in (E). **I** The inverse of the FFT peak frequency defines the oscillation timescale, here plotted versus $V_g$.

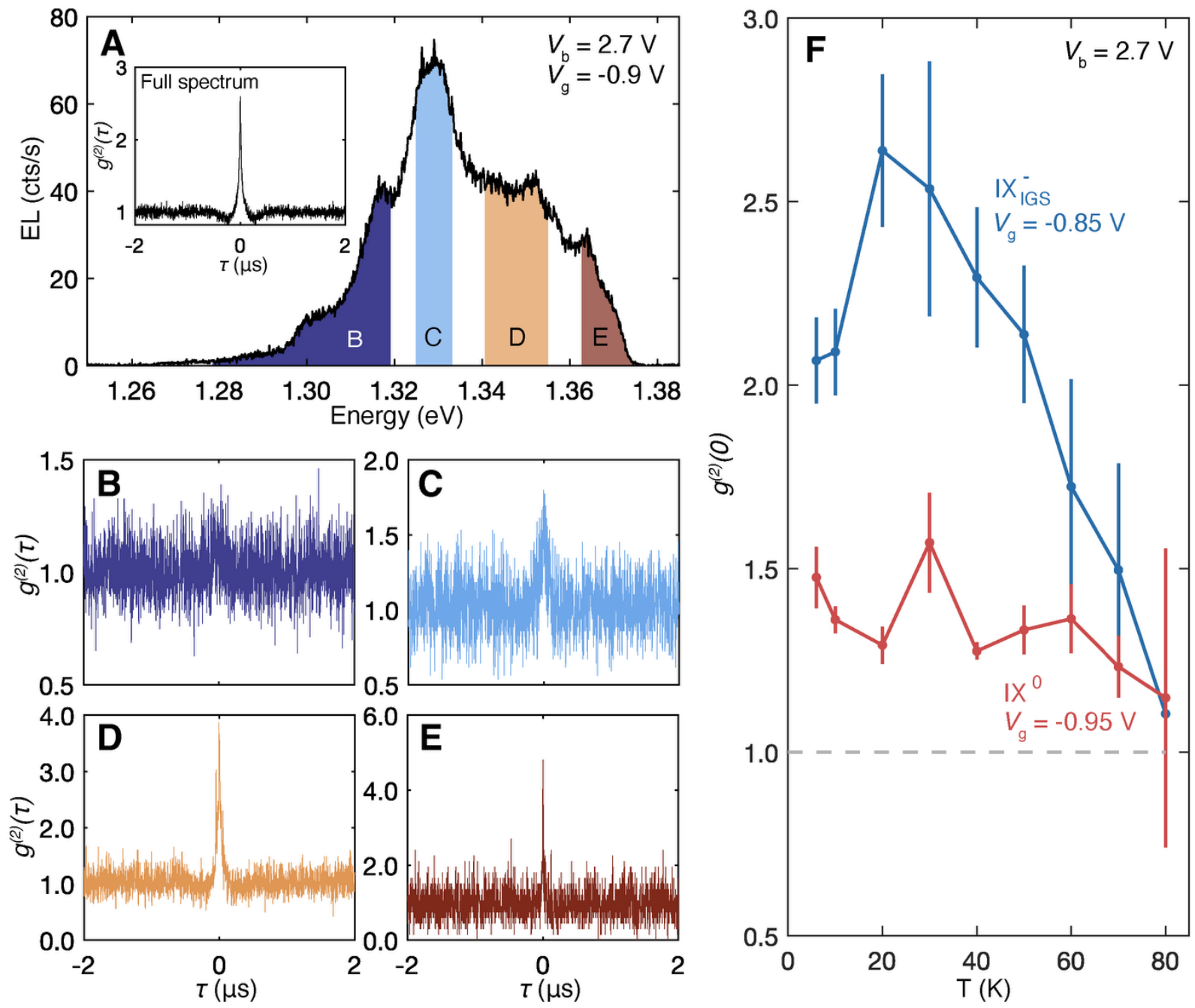

FIG. 3. **Spectroscopic analysis of the origin of the observed photon-photon correlations**. **A** The IX EL spectrum can be separated into different energy bands in the IGS doping regime ($V_b$ = 2.7 V and $V_g = -0.9$ V). The $g^{(2)}(\tau)$ response (inset) originates from higher-energy states, as shown by the $g^{(2)}(\tau)$ traces (**B–E**) obtained from the energy-filtering windows indicated by the colored regions in the EL spectrum. **F** Temperature dependence of $g^{(2)}(0)$ for $V_g = -0.95$ V ($n = p$) and $V_g = -0.85$ V (IGS, $n$-doping), showing that photon-photon correlations persist up to ∼ 80 K.

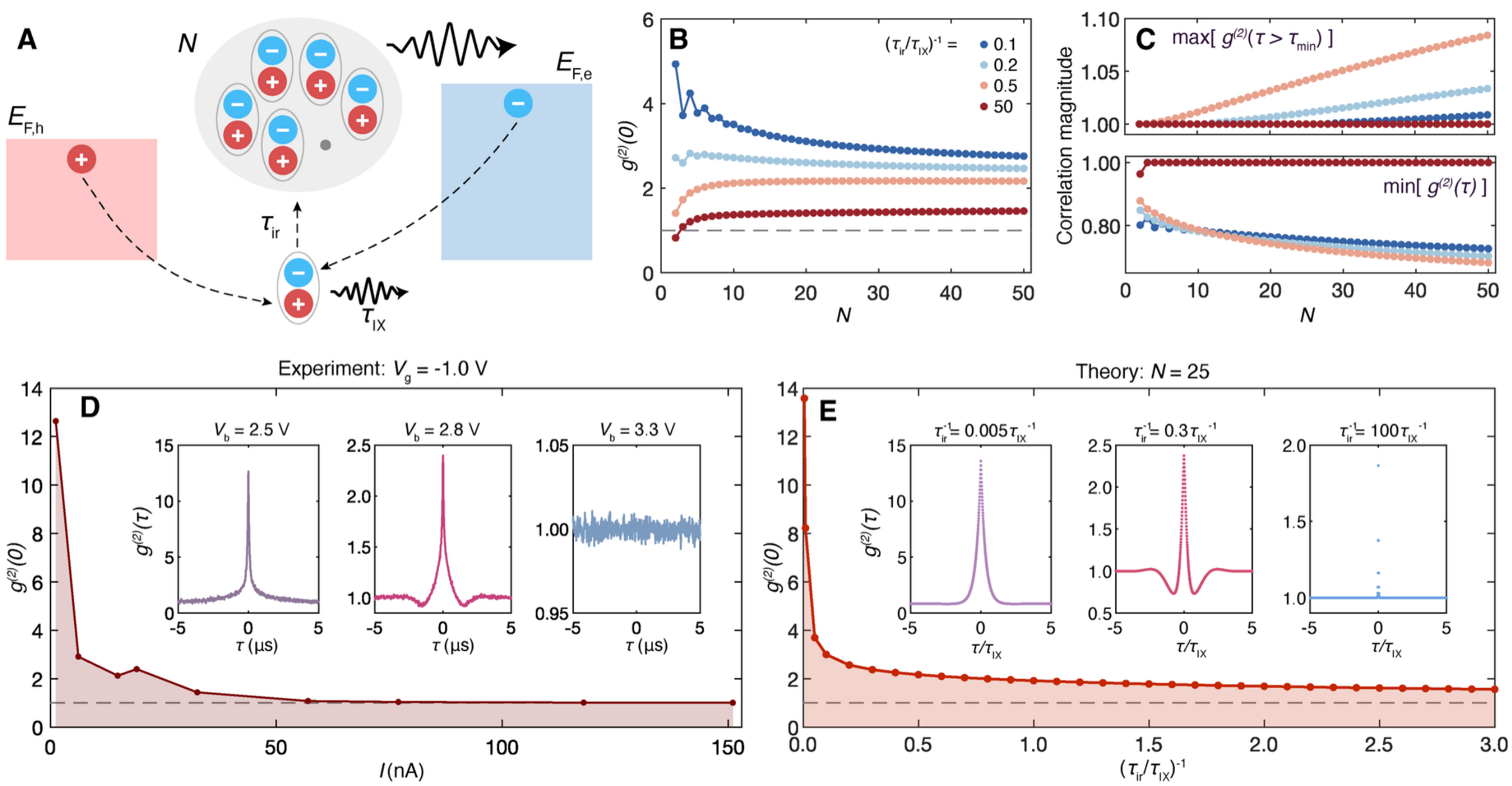

FIG. 4. **Phenomenological model of superradiance with uncorrelated recharging. A** Schematic of the model, where localized emitters are coupled and emit light cooperatively. The traps are incoherently repopulated on a timescale $\tau_{ir}$ by electrically injected electrons and holes that propagate from the reservoirs with Fermi energies $E_{F,e/h}$. Here, $\tau_{IX}$ denotes the radiative lifetime of an isolated emitter. **B** Depending on the model parameters, $g^{(2)}(0)$ can increase (decrease) as a function of the total number of emitters $N$ for large (small) relative recharging rate $(\tau_{ir}/\tau_{IX})^{-1}$. **C** The modeled time-dependent correlation function $g^{(2)}(\tau)$ quantitatively reproduces the key features of the measured behavior, including photon bunching at $\tau = 0$ (B), the subsequent decay and minimum (lower panel), and the revived finite-time photon bunching that follows (upper panel). In (D) and (E), we present a side-by-side comparison between experiment and model. **D** Experimentally measured $g^{(2)}(0)$ as a function of current $I$ at fixed gate voltage $V_g = -1.0$ V. $I$ serves as a proxy for the relative rate $(\tau_{ir}/\tau_{IX})^{-1}$. Insets show $g^{(2)}(\tau)$ traces at selected bias voltages $V_b$. **E** Numerically computed $g^{(2)}(0)$ as a function of $(\tau_{ir}/\tau_{IX})^{-1}$ at a fixed total number of emitters $N = 25$. Insets show $g^{(2)}(\tau)$ traces for representative values of $(\tau_{ir}/\tau_{IX})^{-1}$.